# ARTICLE

# Thermoelectric power factor of pure and doped ZnSb via DFT based defect calculations

Alexandre Berche[a] and Philippe Jund*,[a]

The power factor of pure p-type ZnSb has been calculated via *ab initio* simulations assuming that the carrier concentrations are due to the doping effect of intrinsic zinc vacancies. With a vacancy concentration close to the experimental solubility limit we were able to perfectly reproduce the Power Factor measured in polycrystalline ZnSb samples. The methodology has then been successfully extended for predicting the effect of extrinsic doping elements on the thermoelectric properties of ZnSb. Germanium and tin seem to be promising p-type doping elements. In addition, we give, for the first time, an explanation of why it is difficult to synthesize polycrystalline n-type ZnSb samples. Indeed, compensative effects between intrinsic defects (zinc vacancies) and doping elements (Ga, or In) explain the existence of an optimal (and relatively high) dopant concentration necessary to convert ZnSb into an n-type semiconductor.

## A Introduction

A common theory about semiconductors explains that charged intrinsic point defects provide the carrier concentration necessary to fix the n- or p-type conductivity of the non-doped phase[1]. Based on this fact, it is then possible to calculate the carrier concentration (and afterwards the thermoelectric properties) by combining the calculated formation energy of a given defect with the electronic density of states of the phase containing this defect. Such a methodology has already been applied to the half-Heusler structure NiTiSn where interstitial $Ni_i$ defects allow to reproduce the thermoelectric properties of the non-doped NiTiSn phase[2]. Similarly, for zinc antimonide, it has been shown that the p-type behavior of the compound is due to the presence of zinc vacancies[3,4].

Such a computational procedure can be useful to diminish the cost and time-consuming experimental trial & error methods for searching the best doping elements. Indeed, the efficiency of a thermoelectric module is proportional to the figure of merit ZT (where Z = PF/κ) of the n- and p-type legs constituting the module. Doping a thermoelectric material will mainly improve the electrical transport properties such as the Seebeck coefficient (S) and the electrical conductivity (σ) and as a consequence the power factor (PF = S²σ). The last parameter to determine ZT is the thermal conductivity (κ = $κ_e$ + $κ_l$) where the main contributing part *i.e.* the lattice part ($κ_l$) is mainly due to the morphology of the sample (size of the grains, presence of inclusions…) whereas the electronic part ($κ_e$) is also influenced by doping effects.

The aim of the present paper is to predict the doping effect of zinc vacancies and of foreign elements on the electronic part (PF) of the thermoelectric properties of ZnSb using the *ab initio* methodology mentioned above. However, to be as accurate as possible, calculations have to be done with functionals allowing to reproduce the main properties of a thermoelectric material: structural properties (cell parameters and angles), mechanical properties (elastic constants), electronic properties (band gap and effective masses of the carriers) and the thermodynamic properties (enthalpy of formation of the phases).

In part B, the details of the calculations will be given; section C contains the results on the numerical determination of the PF of pure ZnSb and in part D, doping effects on the PF of ZnSb are tackled. Finally, the major conclusions are drawn in section E.

## B Details of the calculations

The DFT calculations are performed using the Vienna *Ab initio* Simulation Package (VASP)[5,6] and the Projector Augmented Waves (PAW) technique[7,8] within the Local Density Approximation (LDA) or the Generalized Gradient Approximation (GGA). The Perdew – Burke - Ernzerhof parameterization (PBE) is applied[9,10]. In addition, a GGA+U method is used for the d-orbitals of the zinc atoms for which the procedure of Dudarev *et al.*[11] is used (the choice of the parameter Ueff (5eV) is detailed in the Supplementary Data A). Previous studies on ZnSb have shown that both LDA and GGA underestimate the electronic band gap[3,12]. To avoid this problem, meta-GGA or hybrid functionals can be used[12,13,14]. In this study, two meta-GGA functionals: mBJ[15,16] and SCAN[17] have been tested.

Standard versions of the PAW potentials for Sb and Zn are used. The pseudo-potential names are respectively Sb and Zn. Five electronic states are included in the valence shell for Sb ($5s^2 5p^3$) and twelve are taken for Zn ($3d^{10} 4s^2$). For doping elements, the standard potentials are used: Si ($3s^2 3p^2$), Ge ($4s^2 4p^2$), Sn ($5s^2 5p^2$), Pb ($6s^2 6p^2$), Ga ($4s^2 4p^1$) and In ($5s^2 5p^1$).

[a.] ICGM-Université de Montpellier, CNRS, ENSCM, UMR 5253, Montpellier, France.

The calculations are performed using the "accurate" precision setting in the VASP input file to avoid wrap-around errors. The first Brillouin zone is integrated using Monkhorst-Pack k-point meshes. The reciprocal space mesh is set so as to obtain a number of k-points in the irreducible part of the Brillouin zone multiplied by the number of atoms higher than 500. The cutoff energy is set to 500 eV for the whole study. Since the ZnSb phase is paramagnetic (even with defects or dopants), spin-polarization was not taken into consideration.

The calculated cell parameters are obtained by minimizing the total energy of the conventional cell (starting from the experimental structure[18]). Both cell parameters and positions of the atoms have been relaxed. The procedure stops when a difference in energy of 10 μeV or a difference in force below 0.1 meV.Å$^{-1}$ is obtained.

The electronic transport properties are analyzed by solving the Boltzmann's equations using the BoltzTraP code[19] (version 1.2.5) under the constant relaxation time approximation. The first limitation in this theory consists in the usual use of experimental carrier concentrations (N). To avoid this, we will assume that the thermoelectric properties of the material are due to electrons (or holes) provided by the main intrinsic defects (for non-doped materials) or by the doping elements in the case of extrinsic doping. The carrier concentration N can then be estimated using either the densities of states (equation (1)) or the thermodynamic carrier concentrations due to the charged defects (equation (2)). In these equations V is the volume of the supercell, $n_h(\mu_e,T)$ and $n_e(\mu_e,T)$ are the number of holes and electrons respectively in the supercell as defined in equations (3) and (4), q is the charge (in number of electrons), $\mu_e$ is the chemical potential of the electrons and $n_D(\mu_e,T)$ is the number of defects D per supercell. This last term is defined by equation (5) where $N_{site}$ is the number of defect sites per cell of the crystal, $N_{sym}$ the number of symmetrically equivalent ways of introducing the defect on one defect site ($N_{sym}$ = 1 for defects involving one atom such as vacancies or atomic substitutions), $\Delta_{def}E_{charged}$ is the formation energy of the charged defect, $k_B$ the Boltzmann constant and T is the temperature.

$$N = \frac{1}{V}\big(n_h(\mu_e,T) - n_e(\mu_e,T)\big) \quad (1)$$

$$N = -\frac{1}{V}\sum_{q,D} q n_D(\mu_e,T) \quad (2)$$

$$n_h(\mu_e,T) = \int_{-\infty}^{\varepsilon_{VBM}} n(\varepsilon)\big(1 - f(\varepsilon,\mu_e,T)\big)d\varepsilon \quad (3)$$

$$n_e(\mu_e,T) = \int_{\varepsilon_{VBM}}^{\infty} n(\varepsilon) f(\varepsilon,\mu_e,T) d\varepsilon \quad (4)$$

$$n_D(\mu_e,T) = N_{site} N_{sym} \exp\left(-\frac{\Delta_{def}E_{multi}(\mu_e)}{k_B T}\right) \quad (5)$$

To estimate the carrier concentration, we thus need to calculate the formation energy of defects taking into account their charge ($\Delta_{def}E_{multi}$). This methodology and the difference with the one of Zhang and Northrup[1] has been detailed in a previous paper[2]. For a defective cell, the global composition of the cell is generally changed. The energy of the cell has then to be compared to the one of the multi-phased region in equilibrium at the chemical composition of the defective cell. The energy of formation of the defect can then be calculated from equation (6):

$$\Delta_{def}E_{multi}(\mu_e) = \frac{\Delta_f E_{defect}(\mu_e) - \Delta_f E_{multi}}{x_{defect}} \quad (6)$$

where $x_{defect}$ is the concentration of defects in the cell and $\Delta_f E_{multi}$ is the energy of formation of the multi-phased region corresponding to the exact composition of the defective cell. This term is given by equation (7) where φ are the phases involved in the multi-phased region, $x_\varphi$ the volumic fraction of each phase and $\Delta_f E(\varphi)$ the energy of formation of each phase. This last term is given by equation (8) where E(φ) is the DFT-calculated energy of the phase φ; E(M) is the DFT-calculated energy of each constituting element in its standard crystallographic structure and $x_M$ is the atomic fraction of the element M in phase φ.

$$\Delta_f E_{multi} = \sum_\varphi x_\varphi \Delta_f E(\varphi) \quad (7)$$

$$\Delta_f E(\varphi) = E(\varphi) - \sum_M x_M E(M) \quad (8)$$

The first term in equation (6) is $\Delta_f E_{defect}(\mu_e)$ which is the energy of formation of the phase containing the charged defect. This energy is given by equation (9) where $N_M$ is the number of atoms M in the defective structure and E(M) is the DFT-calculated total energy of element M in its standard crystallographic structure. The total energy of the cell containing the charged defect has to be corrected $E_{def}^{corrected}(\mu_e)$ for each charge q (in number of electrons) and is given by equation (10) where $\varepsilon_{VBM}$ is the maximum of the valence band of the pure cell, $\mu_e$ is the chemical potential of the electrons. In this work, an additional correction term is taken into account: the potential alignment ΔV (as defined by Taylor and Bruneval[22]) which allows to refer the charged supercell to the pure supercell.

$$\Delta_f E_{defect}(\mu_e) = \frac{E_{def}^{corrected}(\mu_e) - \sum_M N_M E(M)}{\sum_M N_M} \quad (9)$$

$$E_{def}^{corrected}(\mu_e) = E_{def} + q(\varepsilon_{VBM} + \Delta V + \mu_e) \quad (10)$$

The additional term ΔV is given by equation (11) where $\langle v_{KS}^{bulk}\rangle$ and $\langle v_{KS}^{defect}\rangle$ are the Kohn-Sham potentials of the pure and charged cell respectively.

$$\Delta V = \langle v_{KS}^{bulk}\rangle - \langle v_{KS}^{defect}\rangle \quad (11)$$

Other corrective terms can be taken into consideration. Indeed, due to periodic boundary conditions, the charged point defect could interact with its own image. To correct this effect, several terms have been used[20-26] without any consensus and several codes (such as sxdefectalign[27], PyDef[29] or PyCDT[29]) have been developed. The most common correction consists in adding a Madelung potential energy. However, for most thermoelectric materials, this term tends to over-correct the formation energy of the defect[30] especially for high values of the charge. As a consequence, in this study, no other corrective term has been added.

## C Thermoelectric properties of pure ZnSb

In order to determine the best theoretical description, different DFT functionals have been used to calculate the cell parameters, elastic constants, electronic band gap, hole effective mass and formation energy of pure ZnSb. The results are compared to the literature in the Supplementary Data B. No functional allows to represent all the parameters or properties correctly. However, an overall agreement is obtained for the GGA+U description (+mBJ for the electronic properties) and the SCAN functionnal. Nevertheless, it is not possible to select un-ambiguously the best functional. Therefore, in the rest of this study, the GGA+U and SCAN descriptions will be used to calculate the thermoelectric properties of pure ZnSb

In our approach, we assume that the intrinsic charged defects provide the main charge carriers explaining the n or p-type conductivity of the non-doped phase. In our previous work we have shown that the non-charged zinc vacancy ($VA_{Zn}$) is the most probable defect in ZnSb[3] as also shown in other studies[4]. To study the effect of charged vacancies on the properties of ZnSb, a Zn atom is removed from a conventional 2*2*2 supercell (containing initially 128 atoms) leading to an atomic concentration of vacancies of 0.79%.

### A Formation energy of zinc vacancies and associated carrier concentration.

The energy of formation of $VA_{Zn}$ has been calculated taking into consideration different charged states as a function of the chemical potential $\mu_e$ which varies within or in the vicinity of the band gap (figure 1). Whatever the functional, the lowest formation energy of the zinc vacancy is obtained for a charge q=-2 close to the valence band. This implies that zinc vacancies attract electrons from the ZnSb network leading to the well-known p-type conductivity of the pure material observed in all the experimental measurements[31-37]. This result is similar to the one previously published by Bjerg et al.[4]. Moreover, at $\mu_e$=0eV, the energy of formation of the defect is calculated for q=-2 at 0.25 eV in GGA+U and 0.4eV in SCAN surrounding the 0.3eV calculated by Bjerg et al.[4] in GGA. It is necessary to mention that in their work, Bjerg et al.[4] did not consider the potential alignment term ΔV but took into consideration the Madelung term. In addition, they performed a study on the influence of the size of the cell (varying from 16 to 256 atoms) and their results are extrapolated values for a single defect inside an infinite cell.

To calculate the carrier concentration due to the presence of charged $VA_{Zn}$ using equation (1) or (2), we need first to determinate the value of the chemical potential as a function of temperature. At each temperature, the value of $\mu_e$ is defined in a self-consistent way by imposing the equality of the two expressions of N given in equations (1) and (2).

For each functional, $\mu_e$ is calculated (figure 2a) below half of the calculated band-gap (and even in the valence band for GGA+U) at 300K and its value decreases when temperature increases which is typical of a p-type material. Moreover, whatever the temperature, $\mu_e$ remains in the energy region where q=-2 is the most probable charge (figure 1).

Once the chemical potential is known, the evolution of the hole concentration as a function of temperature can be plotted for the two functionals (figure 2b). Both functionals give a correct representation of the experimental values[32-34]. Especially, our calculations give a good approximation of the highest values of the carrier concentrations measured for polycrystalline ZnSb samples (labelled "Poly" in figure 2b). It is worth noting that for a single crystal (labelled "Single" in figure 2b), the carrier concentrations are significantly lower especially at low temperature.

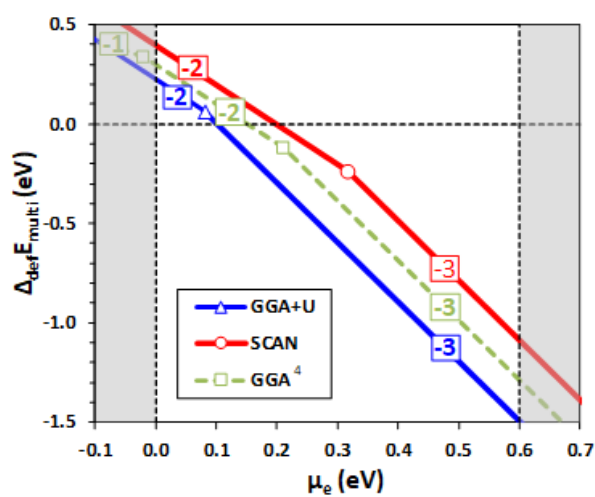

**Fig. 1** Evolution of the formation energy of a charged zinc vacancy with the electronic chemical potential for different values of q (SCAN & GGA+U) compared to the values of Bjerg et al.[4] (GGA).

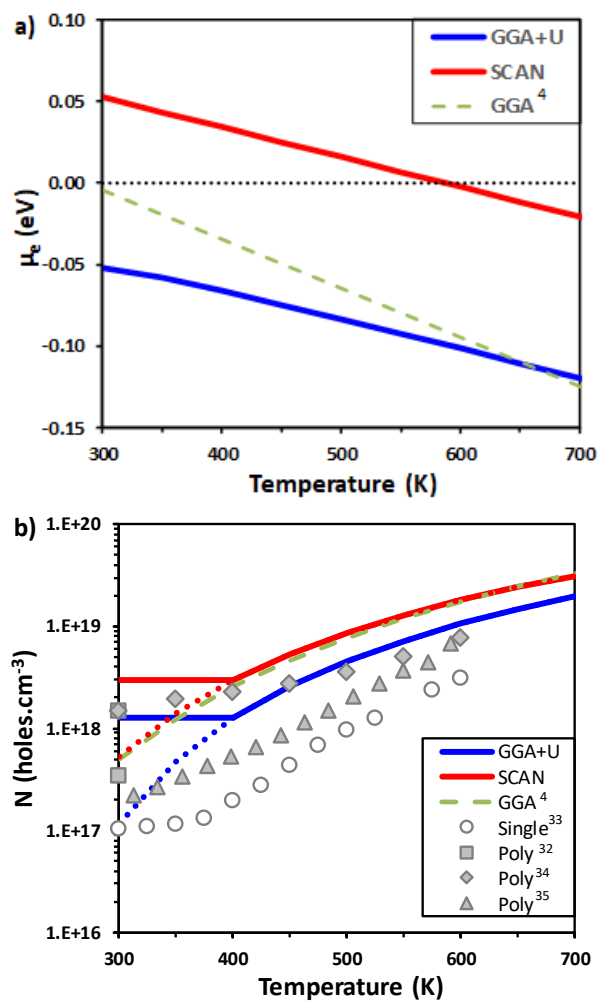

**Fig. 2** Evolution with the temperature of: a) the chemical potential of the electrons; b) the carrier concentration due to $VA_{Zn}$ for different functionals; dotted lines assume thermodynamic equilibrium, solid lines mimic the experimental frozen carrier concentrations below 400K.

It is obvious that these predicted calculated carrier concentrations are given at thermodynamic equilibrium (dotted lines in figure 2 b). However, in experiments, at low temperature (below 400K in this system), this may not be the case leading to almost constant carrier concentrations[33,34]. This fact will be taken into consideration in the following sections of the paper and the calculated carrier concentrations will be artificially frozen below 400K (solid lines in figure 2 b).

**B Seebeck coefficient.**

The Seebeck coefficient has been calculated with the BoltzTraP software[19] using the calculated values of the carrier concentration and the band structures of the solid solution calculated with the GGA+U and SCAN descriptions. The results for the Seebeck coefficient are extremely sensitive to the value of the band gap. This is why, at each temperature, the gap has been fixed to the experimental value (equation D, Supplementary Data B) by applying a rigid band shift operator implemented in BoltzTraP. The evolution of the calculated Seebeck coefficient as a function of temperature for the different functionals is compared to experimental measurements in figure 3. For both GGA+U and SCAN, the calculated Seebeck coefficient is in excellent agreement with the experimental values for polycrystalline samples, which is consistent with the results for the carrier concentrations. However, samples with higher Seebeck coefficients are reported in the literature. Especially, a value of 780µV.K$^{-1}$ is obtained for a single crystal by Hettwer $et\ al.$[33]. This sample has a significantly lower carrier concentration (figure 2 b) then the rest of the samples. This shows that, depending on the sample preparation, the zinc vacancy content can change inducing a variation in the carrier concentration and thus in the value of the Seebeck coefficient. This is directly due to the existence of a homogeneity range in the ZnSb phase which has been experimentally observed and modeled using the Calphad method[38]. The estimated limit of solubility was found for a composition around $(Zn_{0.96}VA_{0.04})Sb$ at 780K and $(Zn_{0.99}VA_{0.01})Sb$ at 300K which surround the global composition of our defective supercell $(Zn_{0.98}VA_{0.02})Sb$. This is why our calculations reproduce the Seebeck coefficients of the experimental samples with the lowest Seebeck values. At the opposite, single crystals have a lower defect concentration and thus exhibit a lower carrier concentration and thus a higher Seebeck coefficient, consistently with figures 2 b and 3.

It is worth noting that the maximum of the Seebeck curve observed at 400K in several experimental samples is directly linked to the fact that thermal equilibrium has not been reached as shown in figure 3 by the difference of shape of the Seebeck curves obtained at thermal equilibrium (dotted lines) and the ones obtained with a frozen carrier concentration at low T (solid lines).

Finally, the experimental Seebeck coefficient of a polycrystalline sample is well reproduced by both, the GGA+U and SCAN, descriptions.

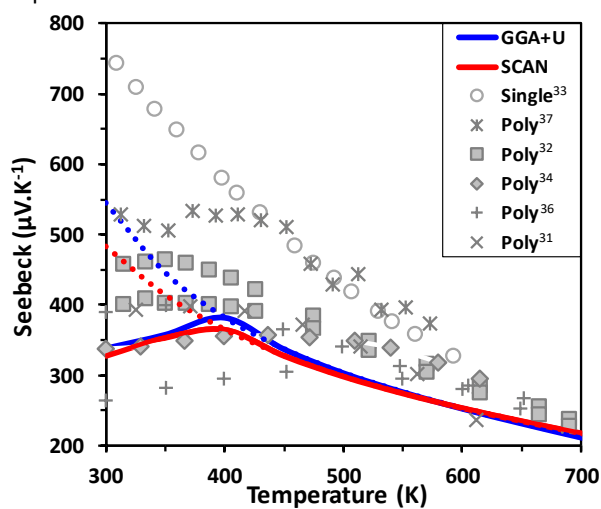

**Fig. 3** Evolution of the Seebeck coefficient as a function of temperature for GGA+U and SCAN compared to experimental data.

**C Electrical conductivity σ and Power Factor (PF).**

Assuming the constant relaxation time approximation, the BoltzTraP code permits to estimate the evolution of σ/τ as a function of temperature, where τ is the electronic relaxation time. To obtain τ the crudest approximation consists in assuming that it does not change with temperature and to fix it at a given value fitted on the experimental electrical conductivity ($10^{-14}$s for ZnSb as suggested by Berland $et\ al.$[34]). This approximation, combined with the hypothesis of the frozen value of N below 400K, allows to perfectly reproduce the experimental values of the electrical conductivity on the whole temperature range (300-700K) (Figure 4) including the plateau at low temperature. Our calculations reproduce the results of the most defective samples, this is why other reported measurements (Böttger $et\ al.$[31] for example) have to correspond to samples with a lower $VA_{Zn}$ content and consequently with a lower carrier concentration and thus a lower electrical conductivity.

As an alternative, to perform a numerical determination of τ one can use the Deformation Potential Theory of Bardeen and Shockley[39]. In this theory, τ depends on the elastic constants in a given direction β ($C_β$) and the effective masses of the holes. However since in ZnSb, $C_β$ and $m_{h,β}^*$ vary tremendously with temperature, it is not possible to obtain a correct evolution of τ as a function of temperature within this theory.

Combining the calculated values of S and σ, the Power Factor has been calculated (Figure 5). An excellent agreement with experimental data is observed demonstrating that the present methodology allows to reproduce well the electronic part of the thermoelectric properties of ZnSb.

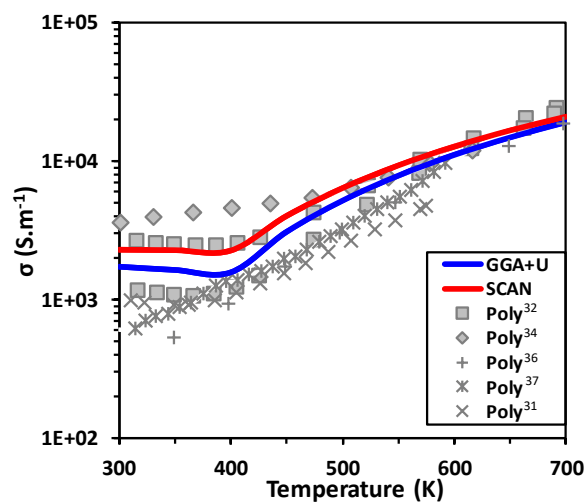

**Fig. 4** Evolution of the electrical conductivity as a function of temperature for GGA+U and SCAN descriptions compared to experimental values

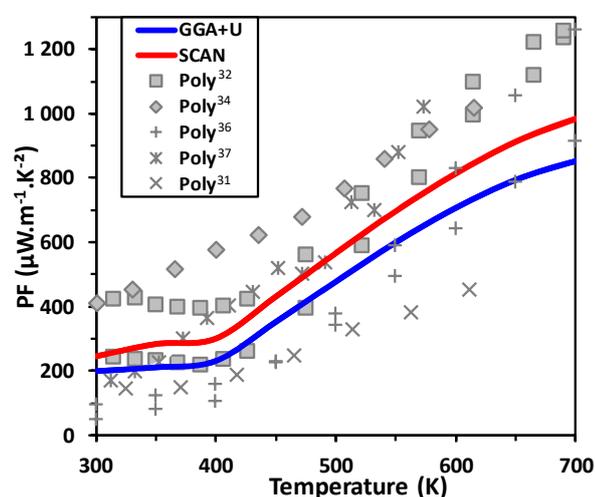

**Fig. 5** Evolution of the power factor (PF) as a function of temperature for GGA+U and SCAN descriptions compared to experimental values.

The BoltzTraP code also permits to calculate the electronic part of the thermal conductivity ($\kappa_e$) divided by τ. This value has to be added to the lattice thermal conductivity ($\kappa_l$) to obtain the total thermal conductivity κ (κ= $\kappa_e+\kappa_l$). To try to obtain a calculated ZT, the previous calculations of Bjerg *et al.*[40] for $\kappa_l$ have been considered. In their work, Bjerg *et al.*[40] considered a mean grain size of 100nm but the calculated lattice thermal conductivity is overestimated by 50%. Therefore, it is not possible to give a correct *ab initio* estimation of ZT here because no other calculation of $\kappa_l$ has been performed since 2014. Since the main aim of this work was to investigate the effect of doping elements on the Power Factor we did not tackle the question of the lattice thermal conductivity of ZnSb in this paper. This is a task on its own especially if a finite displacements method is used in this orthorhombic crystal. In addition, the effect of dopants should be taken into consideration similarly to what has been done in $Fe_2Val$[41], therefore this will be the topic of a forthcoming publication.

## D ZnSb doping

In this last section, we extend the previous methodology to the doping of ZnSb. If numerous p-doping elements (such as Ag, Cu, Sn) are known to improve the figure of merit of ZnSb, it is trickier to synthetize n-type ZnSb. It is possible to obtain temporary n-type ZnSb by Ga, In or Te doping in single crystals, but they turn into p-type after a certain period of time[42]. Explanations for this phenomenon could be the migration of oxygen in the sample[43] or zinc acting as acceptors (similarly to what has been suggested in CdSb[44]). Nevertheless, Ueda *et al.*[35] have shown that it is possible to obtain polycrystalline n-type Te-doped ZnSb with a specific Te content around 2 at. %. Below this concentration, ZnSb is p-type and above this concentration, ZnTe which is known to be a p-type semiconductor, precipitates.

The aim of this section is to validate the extension of our methodology to p-type doping and to understand the difficulty of n-type doping. For each doping element, the expected PF will be calculated. In this section the meta-GGA SCAN description has been selected since we showed in section D that this formalism gives a better reproduction of the experimental PF for "pure" (containing vacancies) ZnSb.

**A p-type doping with Si, Ge, Sn and Pb**

At first, one Si, Ge, Sn or Pb atom has been substituted on a Sb site in a 2x2x2 supercell (corresponding to a concentration of 1.56 at% of the Sb site). The associated formation energy of the defects has been calculated for different charges (Figure 6 a) taking into consideration the ternary phase diagrams described in Supplementary Data C. For these elements, the most stable charge at $\mu_e$ = 0 eV is q=-1 leading to p-type doping as expected. In figure 6 a it can be seen that $Ge_{Sb}$ is more probable than $VA_{Zn}$ whereas for the other dopants, the presence of $VA_{Zn}$ is more probable (or as probable as the presence of the defect). For these last dopants, one has to take into consideration the effects of the dopant together with the zinc vacancies. For that, the formation energy of the two point defects has been calculated (figure 6 b) in a 2x2x2 supercell containing one dopant on an Sb site and one $VA_{Zn}$. The selection of the Sb and the Zn atoms which are substituted has been done using the SQS technique[45,46]. In the presence of two defects, the most stable charge at $\mu_e$ = 0eV is q=-3 which is the sum of the most stable charge of each individual defect.

The carrier concentrations due to the dopant with (for Si, Sn and Pb) or without (for Ge) the zinc vacancy, have been calculated and the predicted TE properties are presented in Figure 7. The results are compared to measurements on polycrystalline samples to have similar zinc vacancy contents. Our calculations allow to give a correct representation of the measurements for Sn-doped compounds[31]. We predict that the Seebeck coefficient of Ge-doped ZnSb should be lower than the one of Sn-doped ZnSb which is consistent with the tendency reported experimentally[47]. For the other elements, no experimental data is available in the literature to our knowledge.

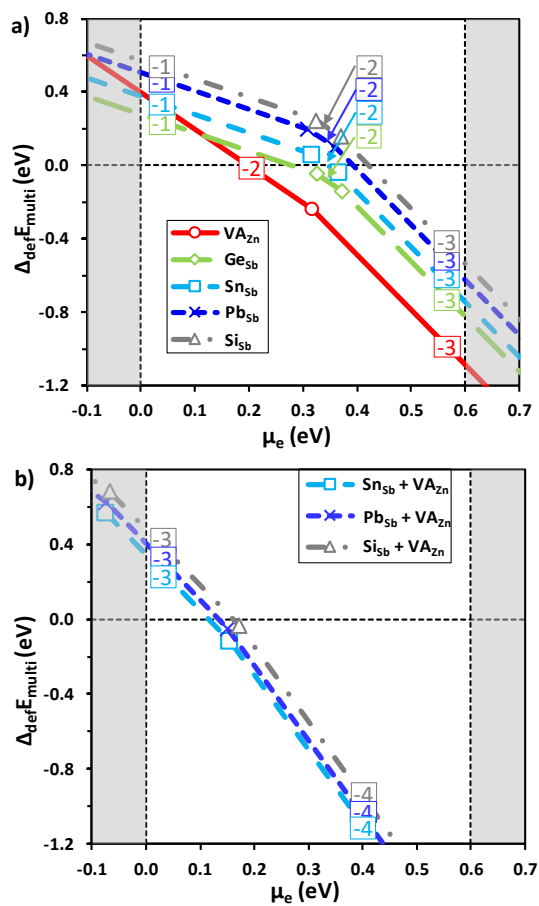

**Fig. 6** Evolution of the formation energy of charged p-type dopants with the electronic chemical potential for different values of the charge q: a) dopant alone; b) dopant with simultaneously one zinc vacancy.

Among these dopants, the best thermoelectric properties are expected for Ge and Sn while Si and Pb do not improve the PF (figure 7 c). With Germanium, the PF is increased by 28% at 700K. If we assume that κ is not changed due to doping (which is certainly false), on the basis of a ZT of 0.8 for pure ZnSb[37], doping with Germanium may lead to a ZT of at least 1 which is close to the maximum value (ZT=1.15) reported in the literature for polycrystalline ZnSb doped with Ag[48]. We can even expect a higher ZT if (as expected) the presence of dopants decreases the value of the thermal conductivity.

**B n-type doping with Ga and In**

For n-type doping, Ga and In have been considered as substituents on the Zn site. The energy of formation of these defects has been calculated for different charges (Figure 8 a) taking into consideration the ternary phase diagrams described in Supplementary Data C. The most stable charge expected at $\mu_e$ = 0 eV is q = 1, confirming the n-type doping. Since all the formation energies are higher than the one of zinc vacancies, $VA_{Zn}$ has to be considered simultaneously. Similarly to what we have observed for p-type dopants, the introduction of one vacancy (q = -2) with one dopant (q = 1) in the supercell, induces a most favorable charge of q = -1 at $\mu_e$ = 0 eV which leads to p-type doping. If one aims to have an n-type compound by Ga or In-doping, the concentration of dopants has to be at least three times more important than the concentration of vacancies. Indeed, at $\mu_e$ = 0 eV, the charge goes from q = -1 for one dopant plus one vacancy to q = +1 for 3 dopants and one vacancy (figure 8 b).

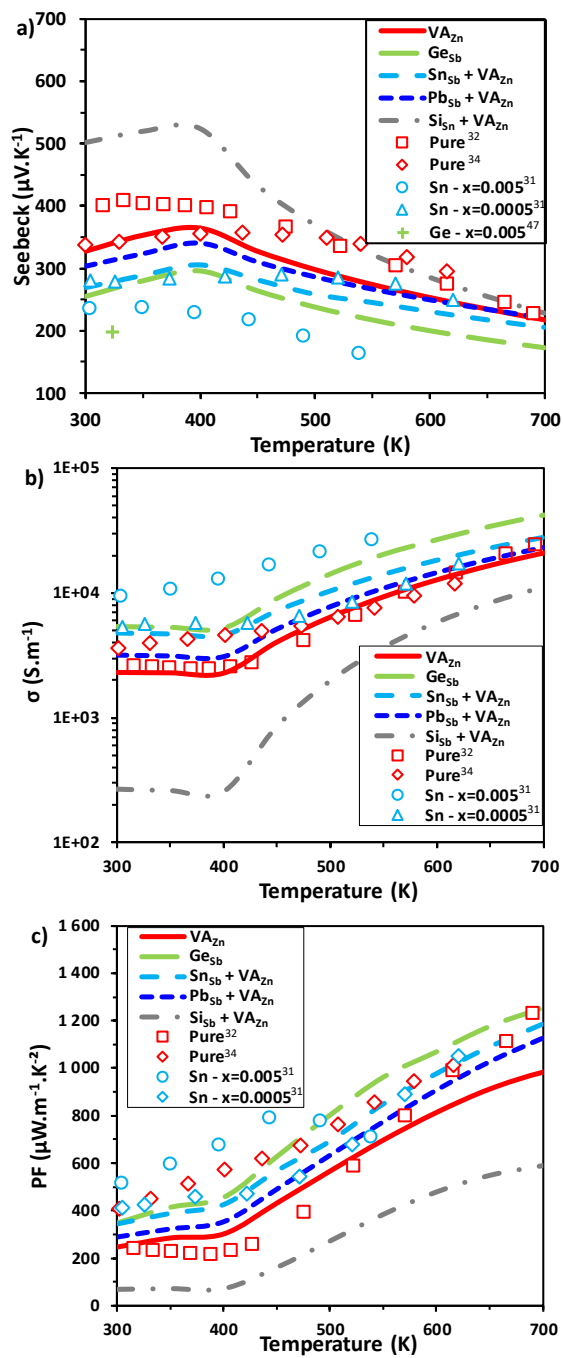

**Fig. 7** Evolution of a) Seebeck coefficient, b) electrical conductivity and c) power factor with the temperature for p-type dopants compared to measurements on polycrystalline samples.

**Journal Name** ARTICLE

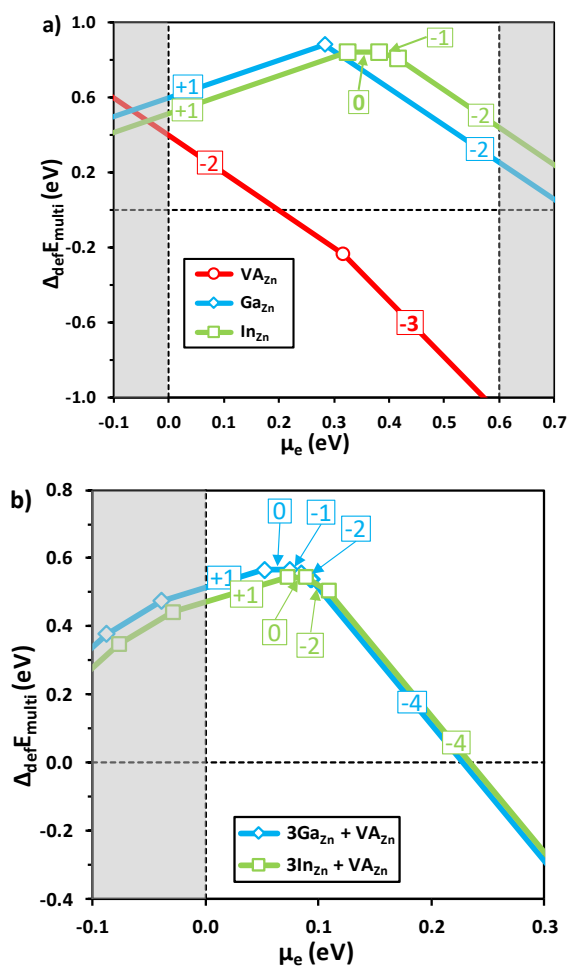

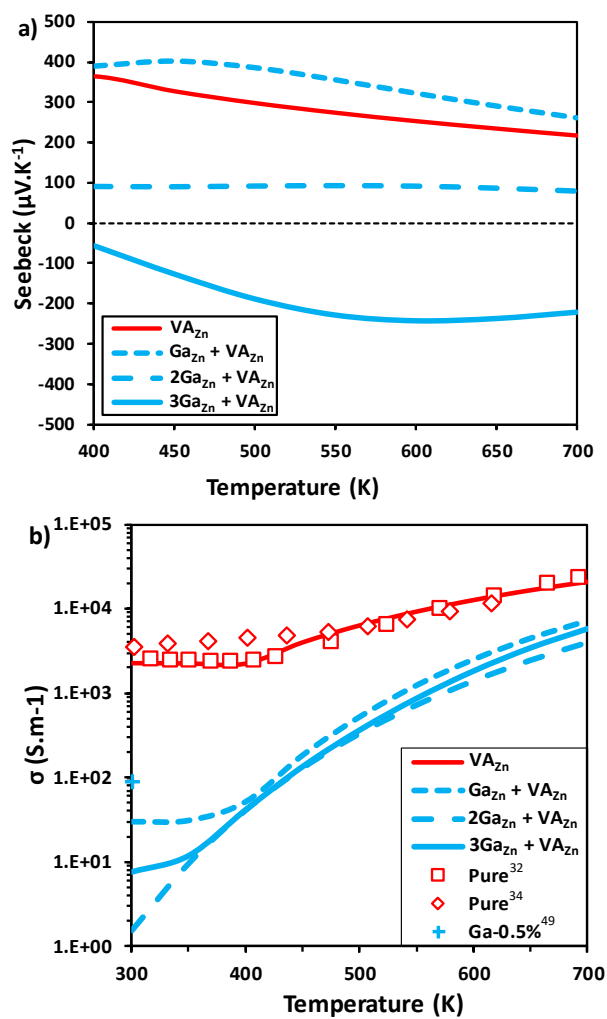

**Fig. 8** Formation energy of defects as a function of the charge q for a) VA$_{Zn}$, Ga$_{Zn}$ and In$_{Zn}$; b) Ga and In doping for 3 dopants plus 1 VA$_{Zn}$ in a 2x2x2 supercell.

We have calculated the Seebeck coefficient and the electrical conductivity of a ZnSb phase doped with different amounts of n-doping elements assuming the presence of one zinc vacancy in a 2*2*2 supercell (for these two quantities, the case of Gallium is shown in figure 9, but the results are similar for Indium (presented in Supplementary Data D)).

For one Ga$_{Zn}$ plus one VA$_{Zn}$ (0.79 at% of Ga in the cell), the carrier concentration is slightly decreased in comparison to one VA$_{Zn}$, but the p-type conductivity remains and the Seebeck coefficient remains almost unchanged. When a second Ga$_{Zn}$ (1.57 at% of Ga in the cell) is added, the most stable charge is q=0 at $\mu_e$ = 0 eV. With such a charge, in our methodology, the carrier concentration is null, however, according to the determination of N (equations (2)), a residual carrier concentration is expected due to other values of the charges and a small Seebeck coefficient is calculated. When the 3rd GaZn (2.36 at% of Ga in the cell) is added in the supercell, an n-type conductivity is predicted (q = +1), electrons become the majority carriers and negative values of the Seebeck coefficient are obtained. This behaviour is summarized in figure 10, where the evolution of the Seebeck coefficient is given as a function of the dopant content in the sample. This figure is given at T=500K since below 400K, it appears that the calculated Seebeck coefficients are anomalously high (in absolute value). This is probably due to the fact that the calculated carrier concentration is small at low temperature (in this case lower than $3.10^{16}$ holes.cm$^{-3}$) and for such small values of N it is well known that the calculated Seebeck coefficient changes enormously with N especially at low temperature. The evolution of the Seebeck coefficient with the dopant content is in agreement with what has been measured by Ueda *et al.*[35] for Te-doped ZnSb. This means that to obtain an n-type ZnSb compound, the concentration of dopants has to be high enough to compensate the concentration of holes due to VA$_{Zn}$ but lower than the limit of solubility of the dopant in the phase. This can explain why it is so difficult to synthesize n-type ZnSb. Indeed, for polycrystalline samples, the doping concentration necessary to have an n-type compound (around 1.5 at% calculated for Ga and In) is higher than the doping concentration generally used for obtaining optimal TE properties and possibly higher than the limit of solubility of Ga and In in ZnSb. At the opposite, the concentration of zinc vacancies being lower in a single crystal, it will be easier to have n-type ZnSb with such samples. This explains why mostly all the n-type ZnSb samples have been reported for single crystals. However, with time, these samples are oxidized and ZnO is formed[32] increasing thus the amount of zinc vacancies in the sample which will lead eventually to an n-type → p-type transition[42].

**Fig. 9** Evolution of: a) the Seebeck coefficient; b) the electrical conductivity as a function of temperature for 2x2x2 supercells containing one zinc vacancy plus zero (solid red), one (blue dashed), two (blue dot-dashed) or three (solid blue) GaZn compared to experimental values[32,34,49].

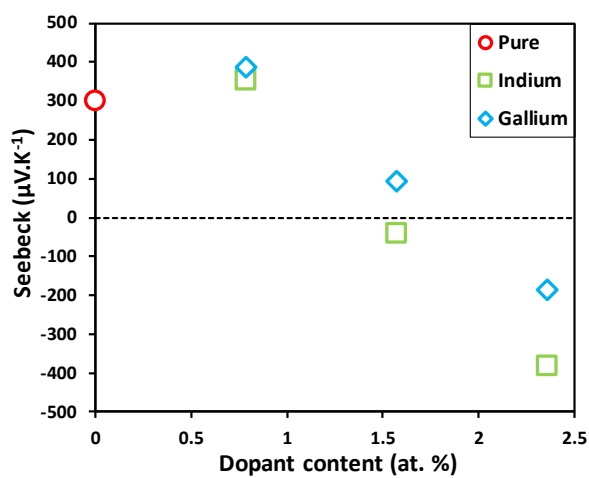

**Fig. 10** Evolution of the calculated Seebeck coefficient as a function of the dopant content in a ZnSb sample at 500K.

Finally, it appears that the carrier concentration of n-type ZnSb will be smaller than the one of p-type ZnSb, leading to a significantly smaller electrical conductivity (figure 9 b). Consequently, the expected PF for n-type doped materials should be significantly smaller compared to the one of p-type ZnSb as shown in figure 11.

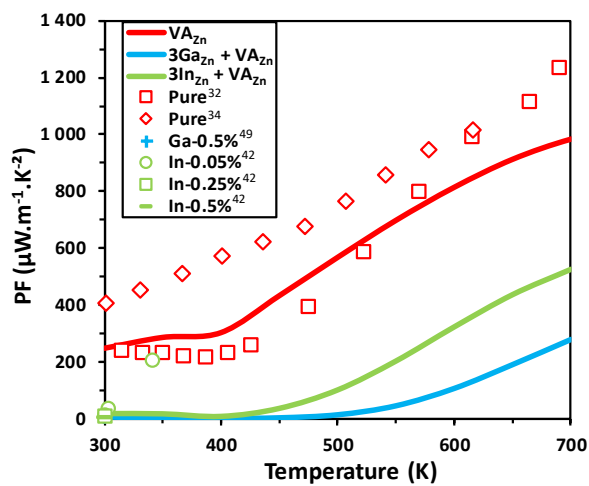

**Fig. 11** Power factor calculated for n-type ZnSb with one VA$_{Zn}$ and 3 Ga or In dopants in the supercell

## E Conclusions

This work confirms the importance of taking into account defects in order to obtain correct calculated thermoelectric properties of a material. As previously shown in NiTiSn[2], the intrinsic defects of ZnSb (Zinc vacancies) are responsible of the experimental concentration of holes, which are at the origin of the p-type conductivity of the pure compound. With the calculated carrier concentration, we are able to predict *ab initio* the Power Factor of the non-doped phase and a good agreement with experiments is obtained showing the quality of the method at least for the electronic part of the ZT. This methodology has then been extended to predict the effect of dopants (n and p-type) on the thermoelectric properties of ZnSb. For p-type doping we predict for Ge (and confirm for Sn) the improved electronic properties. On the other hand, our calculations have shown that for n-type doping, the intrinsic defects have to be taken into consideration because of their counter doping effect. As shown in experiments, this leads then to the existence of a minimum concentration of dopants necessary to change the conductivity type of the host matrix from p to n (or n to p in other cases). On the other hand, the maximum concentration of dopants is given by the solubility limit and if these two limiting concentrations are too close like in ZnSb, it becomes difficult (and sometimes impossible) to synthesize samples with a conductivity type different from the one of the pure compound.

## Conflicts of interest

There are no conflicts to declare.

## References

1  S.B. Zhang and J.E. Northrup, *Phys. Rev. Lett.*, 1991, **67**, 2339
2  A. Berche and P. Jund, *Materials*, 2018, **11**, 868
3  P. Jund, R. Viennois, X. Tao, K. Niedziolka, J.C. Tédenac, *Phys. Rev. B*, 2012, **85**, 224105
4  L. Bjerg, G.K.H. Madsen, B.B. Iversen, *Chem. Mater.*, 2012, **24**, 2111
5  G. Kresse and J. Furthmüller, *Comput. Mater. Sci.*, 1996, **6**, 15
6  G. Kresse and J. Furthmüller, *Phys. Rev. B*, 1996, **54**, 11169
7  P.E. Blöchl, *Phys. Rev. B*, 1994, **50**, 17953
8  G. Kresse and D. Joubert, *Phys. Rev. B*, 1999, **59**, 1758
9  J.P. Perdew, K. Burke, M. Ernzerhof, *Phys. Rev. Lett.*, 1996, **77**, 3865
10 J.P. Perdew, K. Burke and M. Ernzerhof, *Phys. Rev. Lett.*, 1997, **78**, 1396
11 S.L. Dudarev, G.A. Botton, S.Y. Savrasov, C.J. Humphreys and A.P. Sutton, *Phys. Rev. B*, 1998, **57**(3), 1505
12 K. Niedziolka and P. Jund, *J. Elec. Mater.*, 2015, **44**(6), 1540
13 X. He, Y. Fu, D.J. Singh, L. Zhang, *J. Mater. Chem. C*, 2016, **4**, 11305
14 M. Amsler, S. Goedcker, W.G. Zeier, G.J. Snyder, C. Wolverton, L. Chaput, *Chem. Mater.*, 2016, **28**, 2912
15 A.D. Becke and E. R. Johnson, *J. Chem. Phys.*, 2006, **124**, 221101
16 F. Tran and P. Blaha, *Phys. Rev. Lett.*, 2009, **102**, 226401
17 J. Sun, R.C. Remsing, Y. Zhang, Z. Sun, A. Ruzsinszky, H. Peng, Z. Yang, A. Paul, U. Waghmare, X. Wu, M.L. Klein, J.P. Perdew, *Nature Chem.*, 2016, **8**, 831 18   H. Putz and K. Brandenbourg, Pearson's crystal data, Crystal Structure Database for Inorganic Compounds, Cd-Rom software version 1.0., 2007
19   G.K.H. Madsen and D.J. Singh, *Comp. Phys. Comm.*, 2006, **175**, 67
20   S.E. Taylor and F. Bruneval, *Phys. Rev. B*, 2011, **84**, 075155
21   M. Leslie and M. J. Gillan, *J. Phys. C*, 1985, **18**, 973
22   *G. Makov and M. C. Payne*, Phys. Rev. B, 1995, **51**, 4014
23   P.A. Schultz, *Phys. Rev. Lett.*, 2000, **84**(9), 1942
24   C.W.M. Castleton, A. Höglund, S. Mirbt, *Phys. Rev. B*, 2006, **73**, 035215
25   S. Lany and A. Zunger, *Phys. Rev. B*, 2008, **78**, 235104
26   N.D.M. Hine, K. Frensch, W.M.C. Foulkes, M.W. Finnis, *Phys. Rev. B*, 2009, **79**, 024112
27   Freysoldt, J. Neugebauer, C.G. Van de Walle, *Phys. Rev. Lett.*, 2009, **102**, 016402
28   E. Pean, J. Vidal, S. Jobic, C. Latouche, *Chem. Phys. Lett.*, 2017, **671**, 124
29   D. Broberg, B. Medasani, N.E.R. Zimmermann, G. Yu, A. Canning, M. Haranczyk, M. Asta, G. Hautier, *Comp. Phys. Comm.*, 2018, **226**, 165
30   X. Liu, L. Xi, W. Qiu, J. Yang, T. Zhu, X. Zhao, W. Zhang, Adv. Electron. Mater., 2016, 2, 1500284
31   P.H.M. Böttger, G.S. Pomrehn, G.J. Snyder, T.G. Finstad, *Phys. Stat. Solidi A*, 2011, **208**(12), 2753
32   R. Pothin, R.M. Ayral, A. Berche, D. Granier, F. Rouessac, P. Jund, *Chem. Engineering J.*, 2016, **299**, 126
33   K.J. Hettwer, E. Justi, G. Schneider, *Adv. Ener. Conv.*, 1965, **5**, 355
34   K. Berland, X. Song, P.A. Carvalho, C. Persson, T.G. Finstad, O.M. Lovvik, *J. Applied Phys.*, 2016, **119**, 125103
35   T. Ueda, C. Okamura, Y. Noda, K. Hasezaki, *Mater. Trans.*, 2009, **50**(10), 2473
36   Q. Guo and S. Luo, *Func. Mater. Lett.*, 2015, **8**, 1550028
37   C. Okamura, T. Ueda, K. Hasezaki, *Mater. Trans.*, 2010, **51**, 860
38   Y. Liu and J.C. Tédenac, *Calphad*, 2009, **33**, 684
39   J. Bardeen and W. Shockley, *Phys. Rev.*, 1950, **80**,72
40   L. Bjerg, B.B. Iversen, G.K.H. Madsen, *Phys. Rev. B*, 2014, **89**, 024304
41   S. Bandaru, A. Katre, J. Carrete, N. Mingo, P. Jund, *Nanosc. Microsc. Therm.*, 2017, **21**, 237
42   H.G. Müller and G. Schneider, *Z. Naturforsch. A*, 1971, **26**(8), 1316
43   G. Schneider, *Phys. Stat. Solidi B*, 1969, **33**, K133-K136
44   G.A. Silvey, V.J. Lyons, V.J. Silvestri, *J. Electrochem. Soc.*, 1961, **108**(7), 653
45   A. van de Walle, P. Tiwary, M. de Jong, D.L. Olmsted, M. Asta, A. Dick, D. Shin, Y. Wang, L.Q. Chen, Z.K. Liu, *Calphad*, 2013, **42**, 13
46   A. van de Walle, M. Asta, G. Ceder, *Calphad*, 2002, **26**, 539
47   M.I. Fedorov, L.V. Prokofieva, Y.I. Ravich, P.P. Konstantinov, D.A. Pshenay-Severin, A.A. Shabaldin, *Fiz. Tek. Poluprovodnikov*, 2014, **48**(4), 448
48   D.B. Xiong, N.L. Okamoto, H. Inui, *Scripta Mater.*, 2013, **69**(5), 397
49   A. Abou-Zeid and G. Schneider, *Z. Naturforsch. A*, 1975, **30**(3), 381